\documentclass[aps,preprint,showpacs,floatfix]{revtex4}
\usepackage{graphicx,bm,amsmath,dcolumn}
\graphicspath{{figures}}
\usepackage{epsfig}

\begin{document}

\title{The Virial Correction to the Ideal Gas Law: A Primer }

\author{Ron Aaron}
\email{r.aaron$neu.edu}
 \author{F.Y. Wu}
\email{fywu@neu.edu}
\affiliation{Department of Physics, Northeastern University, Boston, MA 02115, USA}

\date{\today}

\begin{abstract}

The virial expansion of a gas is a correction to the ideal gas law that is usually discussed  in advanced courses in statistical mechanics.  In this note we outline this derivation in a manner suitable for advanced undergraduate and introductory graduate classroom presentations.  We introduce a physically meaningful interpretation of the virial expansion that has heretofore escaped attention,  by showing that the virial series is actually
an expansion in a parameter that is the ratio of the effective volume of a molecule to its mean volume. Using this interpretation we show why under normal conditions ordinary gases such as O$_2$ and N$_2$ can be regarded as ideal gases.

  \end{abstract}


\maketitle

 \bigskip
\bigskip
\section{Introduction}
\label{Introduction}
The ideal gas describes a system of $N$ point particles confined in a volume $V$
without interactions and collisions. The ideal gas law
\begin{equation}
 {PV}= {NkT}  \label{idealgas}
\end{equation}
that relates the volume $V$ to the pressure $P$ and the temperature $T$ of the system (with $k$ the Boltzmann constant)
is well-known.  In a typical introductory physics course, the ideal gas law is derived by elementary
considerations in conjunction with the assertion that the average kinetic energy per particle is $3kT/2$. Students
are invariably told without explanation that some real gases, for example N$_2$ and O$_2$, also obey the ideal gas
law,
 and end-of-chapter problems are assigned that involve applying the
law to these gases.
Unfortunately, the implied assertion that the ideal gas law holds in the case of N$_2$ and O$_2$ is misguided and misleading.
In a real gas the mean free path,
the average distance a gas molecule travels between collisions, can be estimated to be of the order
of $10^{-5}$ cm so collisions do occur. This creates an apparent paradox that a gas with inter-molecular interactions
satisfies a relation derived for a gas without interactions.

 In this {\it Note} we address this  paradox. We present a simplified analysis
of the correction to the ideal gas law when there are inter-particle interactions,
 and give the result in a physically meaningful expression
which shows  why the correction can be neglected under usual conditions.
The materials we present is not new and can be found in standard textbooks of advanced statistical mechanics (see, for example, \cite{pathria,mccoy}).  But our discussion is at a level
suitable for instructors of a sophomore level
introductory physics course
who may not necessarily be well versed in the mathematical manipulation in statistical mechanics.

 \section{The virial expansion}
 \label{Virial}
The thermodynamics of a gas system is described by
an equation of state  relating the parameters $T, P, V$ of the system.
The equation of state for a system of gas particles
is formulated and obtained from applications of principles of statistical mechanics.
For a gas system
with  inter-particle interactions described by an internal potential energy in the form of two-particle interactions,
the statistical mechanical analysis is quite complicated involving lengthy mathematical manipulation.
 Therefore,  we first summarize our main result, which is
followed by some details of
the statistical mechanical analysis.

For a gas with two-body
interactions, the analysis using statistical mechanics  leads naturally to
an equation of state in the form of a density expansion,
\begin{equation}
\frac{PV}{NkT} = 1 + B_2 \,\rho + B_3\, \rho^2 + B_4\,\rho^3 + \cdots \label{virialexp}
\end{equation}
where $\rho = N/V $ is the particle density.
The expression (\ref{virialexp}) is known in the literature as the {\it virial expansion}, and
$B_2, B_3,  \cdots$ the second, third, ... virial coefficients.
Explicitly, the virial coefficient $B_k$ is of
the form of a $k$-fold  integral of  a product of a certain
"hole" function $f(r)$ of the order of O(1) 
in a region of ${\bf r}$ of the size of an {\it effective} volume $v_{\rm eff}$ of
a molecule and vanishes for large $r$. Hence $B_k$ is of the order of $v_{\rm eff}^{k-1}$ 
and we can write
\begin{equation}
B_k = c_k \, v_{\rm eff}^{k-1},
\end{equation}
where $c_2, \, c_3,\, \cdots$ are constants (see below). 
 Therefore, (\ref{virialexp})
is in fact an  expansion in   the dimensionless variable
\begin{equation}
\Re = v_{\rm eff}/(V/N) =  v_{\rm eff}/v_{\rm mean} \label{ratio}
\end{equation}
which is the ratio of
the effective volume $v_{\rm eff}$ of a molecule to its mean (per-particle) volume $v_{\rm mean} =  V/N$.
Combining with (\ref{ratio}), 
the virial expansion (\ref{virialexp}) can be written as an expansion in $\Re$ in the form of
\begin{equation}
\frac{PV}{NkT} = 1 + c\,_2 \,\Re + c\,_3 \, \Re ^2 +  c\,_4 \, \Re ^3 + \cdots .\label{virial1}
\end{equation}
Equation (\ref{virial1}) rewrites the virial correction (\ref{virialexp}) to the ideal gas law 
in a physically meaningful
form which appears to have
heretofore escaped attention.
 
Under normal conditions the ratio $\Re$ is very small and in addition,
 because of the convoluted form of the $k$-fold integrals, the coefficients $c\,_k$ also
 diminish as $k$ increases. It follows that the equation of state
(\ref{virial1}) is a rapidly converging series in $\Re$, and reduces
to the ideal gas law (\ref{idealgas}) under normal conditions.

The situation is illustrated by considering the hard sphere gas,
  a gas composed of impenetrable spheres of radius $a$ with no other interactions.
The effective volume of a gas particle is taken to be
the volume of the sphere,
\begin{equation}
v_{\rm eff}   = {4\pi}  a^3/3\, . \label{Veff}
\end{equation}
In this case the integrals in $B_k$ can be exactly evaluated either in closed form or numerically. 
As described in the next section, carrying out the integrals we obtain $c\,_2 =1/2,\  c\,_3 = 5/32, \cdots$
[see (\ref{constantc}) below] 
giving rise to the expansion
\begin{equation}
\frac{PV}{NkT} = 1 + \frac 1 2 \,\Re + \frac 5 {32}\, \Re ^2 +  0.035869\  \Re ^3 + \cdots .\label{virial}
\end{equation}
 For O$_2$, as an example, the ratio $\Re$  can be estimated by
using the hard sphere model with $a = 1.5 \times 10^{-8}$cm\, and\, $v_{\rm mean} 
  =(22,400\ {\rm cm}^3)/(6.023 \times 10^{23}) $, where 
 the denominator is the Avogadro number, or one mole, of molecules which occupy 22,400 cm$^3$ at standard conditions. This yields $\Re = 0.000380$, indicating
 (\ref{virial}) is indeed a fast converging series. For the purpose of
homework in  introductory physics courses, therefore,
 the equation of state (\ref{virial}) of a real gas is well approximated by the deal gas law
(\ref{idealgas}).

\section{The equation of state}
\label{EquationState}
We now return to the analysis of the equation of state with an outline of the statistical mechanics arguments involved. We suggest that
even readers not well versed in the subject will find this section illustrative

In the canonical ensemble formulation of a gas in statistical mechanics, the pressure is given by
\begin{equation}
P = - \bigg( \frac {\partial A} {\partial V} \bigg)_T  \label{pressure}
\end{equation}
where $A = - kT \ln Z_N(V,T) $ is the Helmholtz free energy of the system
 and $Z_N$ is the partition function defined by
\begin{eqnarray}
Z_N(V,T)  &=&  \frac 1 {N!} \int \prod _{i=1}^N (d ^3 {\bf p}_i  d ^3 {\bf r}_i)\ e^{-(K+U)/kT} \label{par} \\
          &=& \frac 1 {N!\lambda ^{3N}} \int \prod _{i=1}^N (d ^3 {\bf r}_i) \ e^{-U/kT}. \label{partition}
\end{eqnarray}
Here  $ {\bf p}_i$  and $ {\bf r}_i$ are respectively  the momentum and coordinate of the $i$-th particle having mass
$m$.  Furthermore,
$K=\sum _{i=1}^N p_i^2/2m$ and $U$ are respectively
the kinetic energy and internal potential energy of the system.  We have also
$\lambda = \sqrt {2 \pi m kT}$
obtained after  carrying out the momentum integrations  in
 (\ref{par}).
 For  the ideal gas $U=0$ and the spatial integration in (\ref{partition}) yields a  simple factor $V^N$.
 Thus we obtain   the ideal gas law (\ref{idealgas}) after substituting (\ref{partition}) into
(\ref{pressure}).

For real gases with inter-molecular interactions, the computation of (\ref{partition})
 is quite complicated. It is commonly assumed that particles interact with pairwise 2-body interactions with
\begin{equation}
U=\sum_{1 \leq i < j \leq N} V(r_{ij})
\end{equation}
where $ V(r_{ij}) = V(|{\bf r}_i  - {\bf r}_j | )= V(r_{ji}) $. 
An example is the Lennard-Jones potential \cite{LJ}
\begin{equation}
V(r) = 4 \epsilon \bigg[\bigg(\frac \sigma r \bigg)^{12} - \bigg(\frac \sigma r \bigg)^{6}\bigg], \label{lennard}
\end{equation}
where $\epsilon$ is the depth of the potential and $\sigma$ the finite distance at which $V(r) = 0$.

In the case of 2-body interactions, the partition function (\ref{partition}) can be analyzed  by using the method of cluster expansion of Mayer and Mayer \cite{mayer}.
However,
while the desired equation of state  is for a fixed $N$, the evaluation of pressure
(\ref{pressure})  is tractable only  in the grand canonical ensemble for which $N$ is not fixed. This
difficulty  is resolved in statistical mechanics
with the introduction of a fugacity $z$ into the grand canonical formulation, with the fugacity
 eventually eliminated  after a long
{\it tour de force} mathematical manipulation to recover $N$. Discussions of this twist
in most text books tend to get entangled in details of the
 algebraic manipulation.  But the end result is surprisingly simple
and is expressed in the physically meaningful
expansion (\ref{virial1}) as we shall see.

The bottom line is that at the end of a lengthy algebraic and combinatoric manipulation (see, for example,
\cite{pathria,mayer}),
  the equation of state emerges  in the form of the   expansion (\ref{virialexp})
 with the  virial coefficient   $B_k$ in the form of a
 $k$-fold integral over a product of the Boltzmann factor $e^{-V(r)/kT}$.  
Specifically, one introduces the hole function
\begin{equation}
f_{ij} = e^{-V(r_{ij})/kT} - 1 
\end{equation}
which is of the order of O$(1)$ in $v_{\rm eff}$ and vanishes for large $r_{ij}$,
one has
\begin{eqnarray}
B_2 &=& -{\frac 1 {2! V}} \int f_{12} d{\bf r}_1 d{\bf r}_2 \nonumber\\
B_3  &=& -{\frac {2} {3!V} }\int \Big[ f_{12} f _{23}f_{31}\Big] d{\bf r}_1 d{\bf r}_2 d{\bf r}_3, \nonumber\\
B_4  &=& -{\frac {3} {4!V} }\int \Big[ f_{12} f _{23}f_{34}f _{41} \big( 3 + 6 f_{13} + f_ {13} f _{24}\big) \Big]
    d{\bf r}_1 d{\bf r}_2 d{\bf r}_3 d{\bf r}_4 \nonumber \\
 \label{B4}
\end{eqnarray}
and generally
\begin{equation}
B_k  = -{\frac {k-1} {k!V} }\int \Big[ {\rm Sum\> of\> irreducible\>\> diagrams \>\>with\>\> } k {\rm \>\>nodes}
\Big] d{\bf r}_1 \cdot\cdot d{\bf r}_k, \label{Bk}
\end{equation}
where we have used a diagrammatic expression for terms in (\ref{Bk}).
Representing product of hole function by lines connecting numbered nodes, the quantity in the square brackets
 is the collection of all {\it irreducible} diagrams of $k$ nodes.
(A diagram is irreducible if it cannot be dissolved  into 2 or more
disconnected pieces with the deletion of a single node, namely, the diagram is
biconnected.) For a description and counting of irreducible diagrams, see \cite{mccoy}.
Particularly, the
diagrammatic representation of $B_2, B_3, B_4$ are shown in Fig. 1. Note the
duplicity factor $3, 6, 1$ in  terms in  (\ref{B4}).

\begin{figure}
\epsfxsize=90mm \vbox to2in{\rule{0pt}{2in}}
\includegraphics{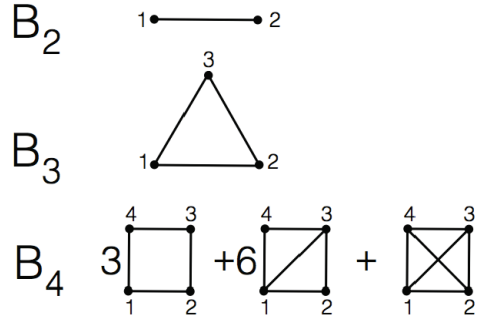}  \caption{Irreducible diagrams representing the integrand in (\ref{B4}). Coefficients $3, 6, 1$ in $B_4$ are multiplicities of equivalent diagrams obtained under the permutation of indices $1,2,3,4$. Line connecting nodes $i$ and $j$ denotes the hole function $f_{ij}$.}\label{fig1}
\end{figure}

The factor $1/V$ in the integrals (\ref{Bk})
is canceled by one of the $k$-fold volume integrations.
For example, in the evaluation of $B_2$ one introduces the change of integration variables $d{\bf r}_1\,d{\bf r}_2
= d{\bf r}_1 \, d{\bf r}_{12}$ and carries out the integration $\int d{\bf r}_1 = V$ to obtain 
\begin{equation}
B_2 = \frac {-1} 2 \int f(r) d{\bf r}  \nonumber
\end{equation}
which is of the order of $v_{\rm eff}$ since $f(r)$ is of the order of O(1) only in a region of ${\bf r}$ in 
 $v_{\rm eff}$ and vanishes for large $r$.
For the Lennard-Jones potential (\ref{lennard}),  we can take without loss of generality, 
\begin{equation}
v_{\rm eff} = 4\pi \sigma^3 /3 \qquad ({\rm Lennard-Jones})\, ,
\end{equation}
which is the volume of a sphere of radius $\sigma$. 
Then, 
 numerically this leads to a 
$B_2$ of the order of $v_{\rm eff}$ or   $B_2 = c\, _2 \,v_{\rm eff}$, where $c\, _2 $ is a constant. 
In a similar manner, one of the $k$-fold integrations in $B_k$ yields a factor $V$
and each of the remaining $(k-1)$-fold integrations yields a factor of the order of $v_{\rm eff}$, as
so  we obtain $B_k = c\, _k \,v_{\rm eff}^{k-1}, k=3,4,...$ 
alluded to earlier, where $c\,_k$ is a constant .
The substitution of  $B_k = c\, _k \,v_{\rm eff}^{k-1}$ into (\ref{virialexp}) now gives the virial 
expansion in the form of  (\ref{virial1}).

\begin{figure}
\epsfxsize=90mm \vbox to2.5in{\rule{0pt}{3.5in}}
\includegraphics{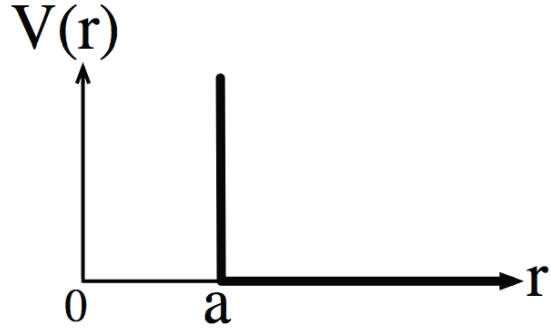}  \caption{The hard sphere potential (\ref{radius}).}
\end{figure}

In the case of a hard sphere gas where molecules are impenetrable spheres of radius $a$, the
 interaction can be represented by the potential $V$ shown in Fig. 2 or, algebraically,
\begin{eqnarray}
V(r) &=& \infty, \hskip 1cm r\leq a \nonumber \\
     &=& 0, \hskip 1.3cm r > a \label{radius}
\end{eqnarray}
with the hole function
 \begin{eqnarray}
f(r) &=& -1, \hskip 1cm r\leq a \nonumber \\
     &=& 0, \hskip 1.3cm r > a. \label{hardspheref}
\end{eqnarray}
The effective volume of a hard sphere gas particle is taken to be the volume of the sphere
\begin{equation}
v_{\rm eff} = 4 \pi a^3/3  \qquad ({\rm hard\>\>sphere}).
\end{equation}

The evaluation of virial coefficients for hard spheres has a long history (for a review see \cite{mccoy}).
$B_2$ is trivially evaluated to be $B_2 = 2\pi a^3/3$, or $c\,_2 = 1/2$. The evaluation of $c\,_3$
 and $B_3$ can be found  in
\cite{pathria,mccoy} and the evaluation of $c\,_4$ and $B_4$ was due to Boltzmann \cite{Boltzmann} and Majumdar
\cite{Majumdar} (see \cite{note1}). The results  are
\begin{eqnarray}
c\,_2 &=& 1/2 = 0.5 ,  \nonumber \\
  c\,_3 &=& 5/32 = 0.152\ 250 , \nonumber \\
  c\,_4 &=&  \frac 1 {2240} \bigg(\frac{219\sqrt 2 + 4131 \tan^{-1}\sqrt 2\  }
 {8 \pi} - {89}  \bigg) \nonumber \\
     &=& 0.035\ 868\ 69, \label{constantc}
\end{eqnarray}
as in (\ref{virial}) or, equivalently, 
 \begin{eqnarray}
B_2 &=& {2 \pi a^3}/3 ,  \nonumber \\
   {B_3}  &=& (5 /8)\, B_2^2  = 0.625\ B_2^2 ,\nonumber \\
  {B_4}    &=& 0.286\ 949\ 51 \ B_2^3   \label{B5}
\end{eqnarray}
as usually given in the literature and standard textbooks \cite{pathria,mccoy}.
 Virial coefficients  up to $B_{10}$ and in higher dimensions are given in \cite{mccoy}.

\section{Summary}
\label{Summary}
We have introduced in Eq. (\ref{virial1}) a simple and concise derivation of the virial correction to the
ideal gas law. The results are presented as an expansion in terms of a parameter
$\Re$ which is the ratio of the effective volume of a molecule to its mean
(per-molecule) volume.  This physically meaningful interpretation of the virial expansion
appears to have heretofore escaped attention.  For real gases, the parameter $\Re$  is extremely  small under usual
conditions and therefore the equation of state effectively
reduces to the ideal gas law. Thus treating gases such as O$_2$ and N$_2$ as an ideal gas
in homework problems is justified.  But more important is the fact that the material we have presented is
not accessible to undergraduate students in an undergraduate textbook.  Yet we feel that it can be understood
and the material considerably broadens their horizon concerning physics and science in general.


\begin{thebibliography}{10}
\bibitem{note1} However, the closed-form expression of $B_4$ cited in \cite{mccoy} is incorrect. We are indebted to B. M. McCoy
for clarifying this point.
\end{thebibliography}
\end{document}